# Experimental Investigation of the Effects of Gas Oil and Benzene on the Geotechnical Properties of Sandy Soils

Faezeh Hanaei[1]†, Mohammad Sina Sarmadi[2]†, Milad Rezaee[3], Aida Rahmani[4]*

1. Faculty of Engineering, Department of Civil Engineering, Kharazmi University, Tehran, Iran
2. Faculty of Engineering, Department of Civil Engineering, Kharazmi University, Tehran, Iran
3. Chemical and Petroleum Engineering Department, Schulich School of Engineering, University of Calgary, Calgary, Canada
4. Center for Infrastructure Engineering, Western Sydney University, Penrith, NSW 2751, Australia

†These authors contributed equally to this work.

**Abstract:**

In recent years, contamination of soils by different hydrocarbon has drawn the attention of geotechnical engineers. However, to our knowledge, no systematic studies have been conducted to investigate the effects of gas oil and benzene contamination on the geotechnical properties of sandy soils. Therefore, the present study investigates the impact of gas oil and benzene contamination on geotechnical properties of two kinds of sandy soil using several laboratory tests such as direct shear, compaction, and permeability tests. The compaction test results showed that optimum moisture content declined in oil-contaminated samples. Otherwise, the maximum dry density increased up to 5% oil content and then decreased by adding oil to the sand soils. The direct shear test indicated that cohesion increased and friction angle decreased after oil contamination. The result of the permeability test revealed that contamination caused a decrease in permeability of sandy soil and the change was a function of oil content and viscosity of oils.

*Corresponding author; rahmani63@gmail.com

## 1. Introduction

Hydrocarbons such as gasoline, diesel fuel, gas oil, crude oil, coal oil, and benzene have attracted the attention of researchers due to their various applications in different industrial sectors. However, their main problem is the threat posed to the environment by leakage from oil spillage, natural deposits, transporting petroleum, corroding pipelines, contamination during production, and separation process. Hydrocarbon contamination, which may occur intentionally or unintentionally, affects the physical and geotechnical properties of the surrounding soil [1]. Shear strength, cohesion, internal friction angle, Atterberg limits, permeability, and consolidation are among parameters that might be changed after contamination [2]. Different methods have been proposed to purify soil from oil compounds which are not economical [3]. The use of oil-contaminated soil in civil projects is a promising and cost- effective solution to reduce its adverse effects on the environment [1].

To date, researchers have carried out different studies to investigate the influences of hydrocarbon contamination on the geotechnical properties of soils. Effects of engine oil, lamp oil, light crude oil, diesel, gasoline, and kerosene on the geotechnical properties of sand have been studied by several researchers [1, 2, 4, 5, 6, 7, 8]. Effect of hydrocarbon contamination on the internal friction angle [1, 2, 4, 5, 6], cohesion [1, 2, 5, 6], compaction behaviour [2, 4, 5, 6], permeability [1, 2, 4, 6], small strain shear modulus [7], and liquefaction potential [9] of sand have been investigated. In addition, effects of hydrocarbon contamination on the bearing capacity of shallow foundations constructed on sandy soils, strip foundations near a slope, and piles in oil-contaminated sand have also been taken into consideration [10, 11, 12, 13].

Previous research studies, those focused on the oil contamination effects on fine-grained soils, particularly clayey soils, investigated the impact of pollution on different geotechnical properties such as the unconfined compressive strength, Atterberg limits, permeability, compressibility, and consolidation behaviours [5, 14, 15, 16, 17]. Moreover, the triaxial shear behaviour of oil-contaminated clay [18] and cyclic behaviour of oil-contaminated clayey soil [19] have been studied.
A review of the above-mentioned literatures shows that although there are many research studies conducted to investigate the effects of different contaminants on the geotechnical behaviour of soils, there is no comparative study on the results of benzene and gas oil contamination effect on the geotechnical properties of sandy soils. Therefore, the aim of this study is to carry out several laboratory tests such as direct shear, compaction, and constant head permeability tests on the clean, benzene- and gas oil-contaminated sand to evaluate the effects of contaminations on geotechnical properties of sandy soil.

## 2. Materials and Methods

### 2.1. Soils

Figure 1 illustrates the particle size distribution curves for two sandy soils used in the present research based on ASTM D6913 [20]. The properties of the used sandy soils are also shown in Table 1. Based on the values of $C_C$ and $C_u$, both samples are poorly graded sand [21,22]. ASTM D854 [23] and ASTM D2434 [24] were used to determine the specific gravity and the hydraulic permeability of the samples, respectively. Proctor compaction tests were also conducted following

ASTM D698 [25] to obtain the maximum dry density and optimum moisture content of samples. Finally, direct shear tests were performed according to ASTM D3080 [26] to measure strength parameters.

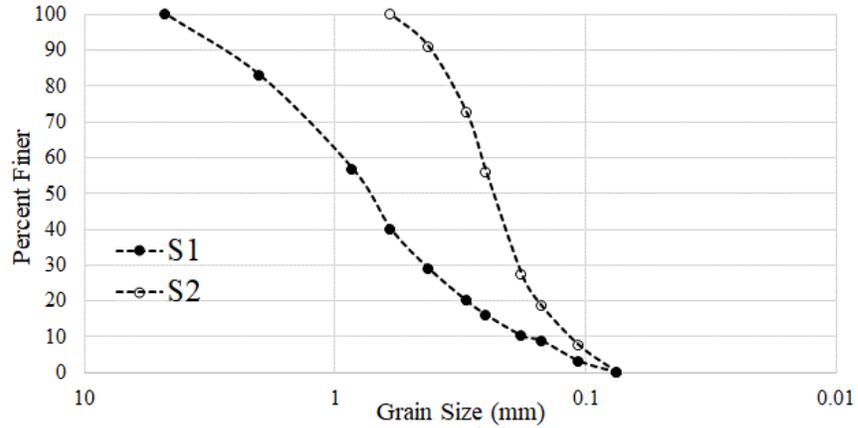

Figure 1. Grain size distribution curves of S1 and S2 sandy soils

Table 1. Properties of sandy soils used in this study [2]

| Characteristics | Results for S1 | Results for S2 |
| --- | --- | --- |
| Specific Gravity ($G_s$) | 2.69 | 2.67 |
| Maximum dry density ($\rho_{dmax}$ $g/cm^3$) | 1.96 | 1.69 |
| Optimum moisture content ($\omega_{opt}$ %) | 13.76 | 13.05 |
| Permeability (cm/s) | 72.5×10⁻³ | 80.8×10⁻³ |
| Void ratio ($e$) | 0.54 | 0.7 |
| $D_{10}$ (mm) | 0.17 | 0.122 |
| $D_{30}$ (mm) | 0.462 | 0.193 |
| $D_{60}$ (mm) | 0.94 | 0.257 |
| $C_C$ | 1.34 | 1.16 |
| $C_u$ | 5.53 | 2.1 |

### 2.2. Benzene and gas oil

The main objective of the present work is to investigate the effects of two different hydrocarbons on the engineering properties of the sandy soils. For this purpose, gas oil and benzene were selected as the target contaminants. The general features of these hydrocarbons, including viscosity and density, are summarized in Table 2.

Table 2. Properties of consumed oils

| Oil Type | Dynamic Viscosity (cp) | Density (kg/l) |
|----------|------------------------|----------------|
| Benzene | 0.61 @ 25 ℃ | 0.882-0.886 @ 15.6 ℃ |
| Gas Oil | 3.68 @ 25 ℃ | 0.82-0.86 @ 15 ℃ |

**2.3. Preparation of Contaminated Soil**

In order to achieve artificially contaminated samples, the soil samples were dried at an oven temperature of 105 ℃ for 18 h and then mixed with different percentages of benzene and gas oil (0, 5, 10, and 15%) by the dry weight of sand. At first, benzene and gas oil were sprayed on the samples and the samples were then mixed manually to have a uniform and homogeneous mixture. Afterward, the samples were put inside a plastic container for 14 days to reach an equilibrium condition. The plastic containers were covered by a lid to prevent the evaporation of gas oil and benzene. The name of the samples is tabulated in Table 3 and 4.

Table 3. Benzene-contaminated Samples

| Sample Name | | Benzene Content (%) |
|-------------|------|---------------------|
| S1 | S2 | |
| S10 | S20 | 0 |
| S1B5 | S2B5 | 5 |
| S1B10 | S2B10 | 10 |
| S1B15 | S2B15 | 15 |

Table 4. Gas oil-contaminated samples

| Sample Name | | Gas Oil Content (%) |
|-------------|------|---------------------|
| S1 | S2 | |
| S10 | S20 | 0 |
| S1G5 | S2G5 | 5 |
| S1G10 | S2G10 | 10 |
| S1G15 | S2G15 | 15 |

## 3. Results and discussions

### 3.1. Compaction Test

The standard Proctor compaction tests were conducted following ASTM D698 [25] and [27]. To calculate the moisture content of hydrocarbon contaminated samples Eq. 1was used [28].

$$\omega \% = (1 + mn)\frac{W_t}{W_d} - (1 + n) \qquad (1)$$

Where $W_t$ is the wet weight of contaminated soil, $W_d$ is the dry weight of contaminated soil, $m$ (%) is the residual content of oil after drying, and $n$ (%) is the oil content before drying. Figures 2 and 3 show the compaction tests results in the form of moisture content versus dry density. As indicated in Figures 2 and 3, the bell-shaped compaction diagram for oil-contaminated samples tended towards the left side of the uncontaminated for both benzene and gas oil-contaminated samples. Therefore, the hydrocarbon-contaminated samples reached their maximum dry density in lower moisture content.

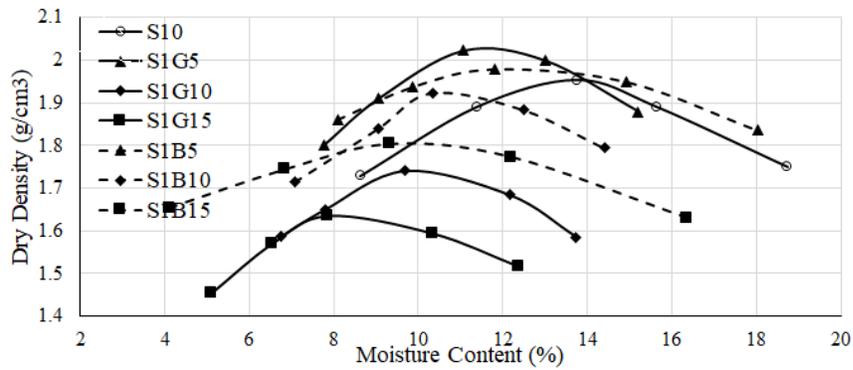

Figure 2. Results of compaction tests for S1 containing different gas oil and benzene contents

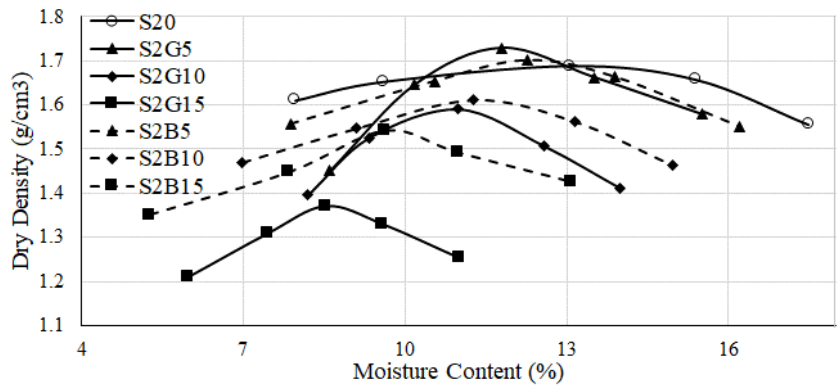

Figure 3. Results of compaction tests for S2 containing different gas oil and benzene contents

According to Figure 4, the maximum dry density for both benzene- and gas oil-contaminated samples increased up to 5% oil content. Then the maximum dry density declined by increasing oil content in soil samples. Increasing maximum dry density in the hydrocarbon-contaminated samples could result from the lubricating effect of oils which facilitates the compaction [29]. For S1 and S2 samples after adding 5% oil, the maximum dry density declined by increasing oil content. By increasing gas oil and benzene contents, the excessive volume of fluid in the sample could disperse the energy of the compaction test and the sample cannot be compacted well enough; accordingly, the maximum dry density declined by increasing oil content. The rate of changes in maximum dry density for gas oil-contaminated samples was higher than that of benzene contaminated samples. This phenomenon might occur because gas oil had a more lubricating effect on the sand than benzene. Similar results have been reported by other researches [4], and it has been observed that hydrocarbon with higher viscosity resulted in a higher rate of changes in maximum dry density. Furthermore, in the current study, the maximum dry density increased followed by a decrease with increasing the oil content. This trend was reported by other researchers. Shin and Das [4] reported an increase in dry unit weight after hydrocarbon contamination. Al-sanad et al. [29] claimed that dry density increased with the presence of oil contamination up to 4%, and for 6% oil contamination the dry density decreased.

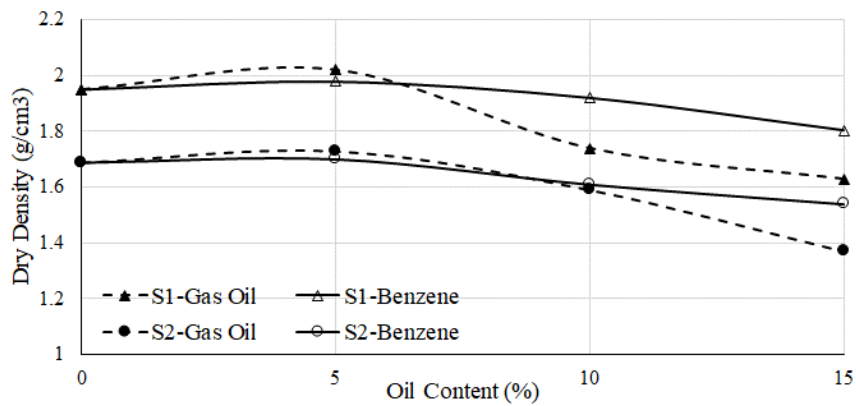

Figure 4. Variation of maximum dry density with gas oil and benzene content

Figure 5 illustrates the influence of benzene and gas oil content on the optimum moisture content. As indicated in Figure 5, an increase in oil content for S1 and S2 samples caused a decline in optimum moisture content, and other researchers have declared similar results for optimum moisture content [4, 5, 28, 29]. As shown in the previous paragraph, gas oil and benzene had a lubricating effect, which could reduce the friction between the sand particles; therefore, by less moisture content, samples could gain their maximum dry density. Moreover, the same as dry density, variation in moisture content is higher for gas oil-contaminated samples compared to benzene contaminated samples. In the same way, the lubricating effect of gas oil due to higher viscosity could be the cause of this difference.

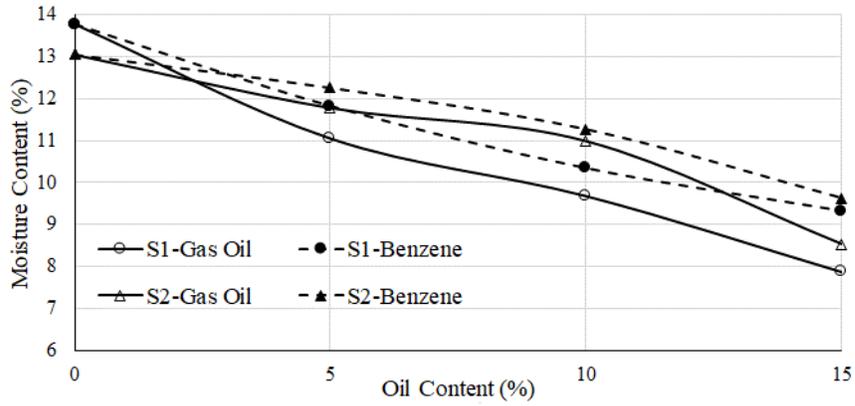

Figure 5. Variation of optimum moisture content with gas oil and benzene content

### 3.2. Direct Shear Test

In the present study, to determine shear strength parameters, the direct shear tests were carried out. Some direct shear tests were conducted on the uncontaminated and hydrocarbon-contaminated samples. The direct shear tests were performed according to ASTM D3080 [26]. The square box was selected with dimensions (10 cm ×10 cm). The direct shear tests were performed with different normal stresses (50, 100, 150, 200, and 250 kPa) under drained conditions. The direct shear tests were carried out under a constant rate of 0.7 mm/min that is proposed by ASTM D3080. Figures 6 and 7 illustrate the normal stress versus shear stress during direct shear tests. Figure 6 indicates the results of direct shear tests on the S1 samples based on the shear stress versus normal stress. In the same way, Figure 7 indicates the results of direct shear tests on the S2 samples. As can be seen, increasing the percentage of oil resulted in the slope of the line to decline and the intercept to raise, which shows a decline in the angle of internal friction and an increment in cohesion.

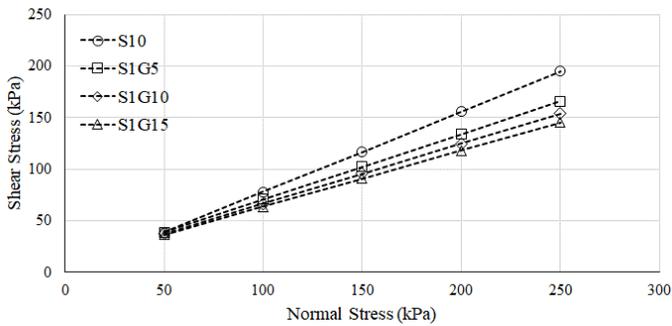

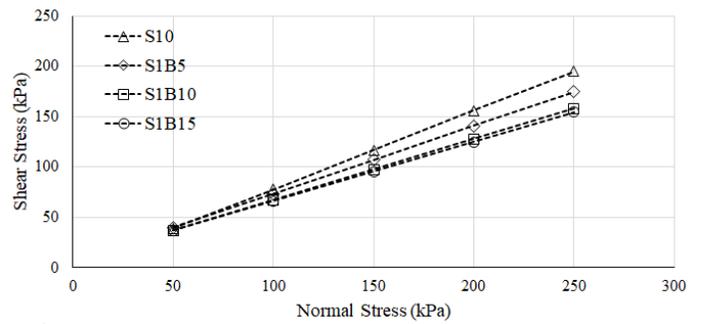

(a)                                                                 (b)

Figure 6. The shear envelope of S1 (a) gas oil-contaminated (b) benzene-contaminated samples

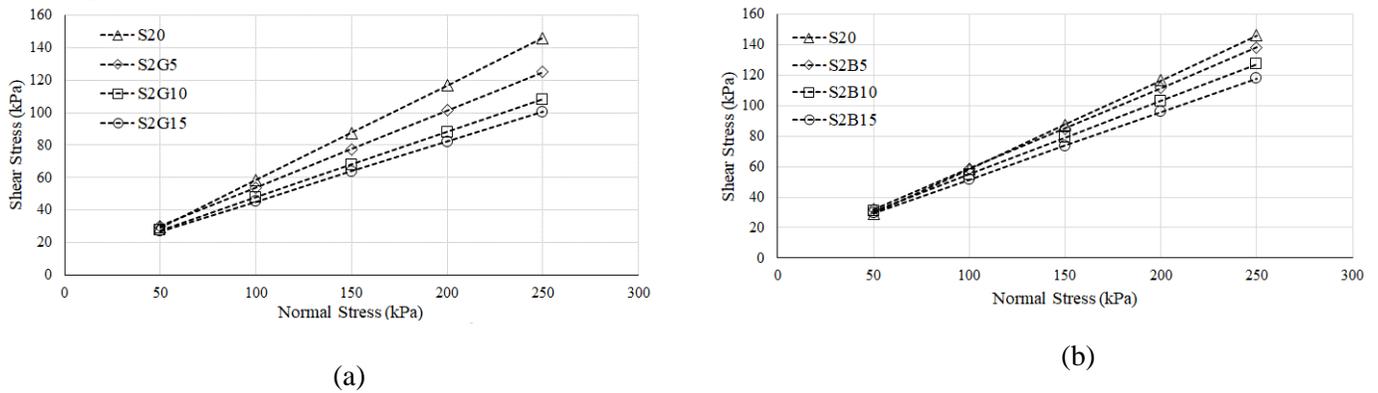

(a)                                                                (b)

Figure 7. The shear envelope of S2 (a) gas oil-contaminated (b) benzene-contaminated samples

As shown in Figure 8, increasing gas oil and benzene contents resulted in a decrease in the angle of internal friction for S1 and S2 samples. The rate of changes in friction angle for gas oil-contaminated samples was higher than that of benzene-contaminated samples. Figure 9 illustrates the influence of gas oil and benzene on cohesion, which indicates that hydrocarbon contamination caused an increment in cohesion for either S1 or S2 samples. Gas oil was more influential than benzene in altering cohesion. Shin and Das [4] performed a set of direct shear tests on oil-contaminated sandy soil. It was reported that samples contaminated with engine oil and had higher viscosity resulted in a higher rate of changes in friction angle and cohesion compared to crude oil and lamp oil with lower viscosity.

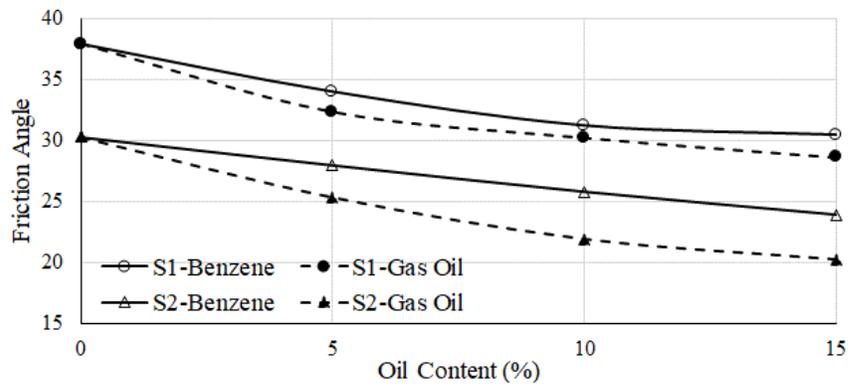

Figure 8. Effect of gas oil and benzene on friction angle of S1 and S2 samples

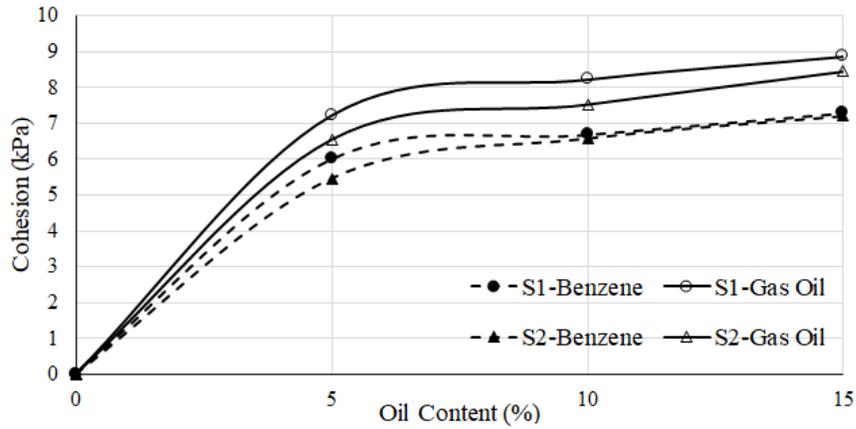

Figure 9. Effect of gas oil and benzene on the cohesion of S1 and S2 samples

### 3.3. Constant Head Permeability Test

Figure 10 shows the effects of gas oil and benzene on the hydraulic permeability of S1 and S2 samples. According to Figure 10, hydrocarbon contamination resulted in a decrease in soil permeability. The hydraulic permeability of S1 and S2 samples decreased after contamination. Hydraulic permeability reduction may be because gas oil and benzene occupy the pore space of the soils, and fluid flow would be easy. The decline in permeability for contaminated soils has been reported by others [4, 28, 29]. Furthermore, the rate of reduction for the gas oil-contaminated sample was higher than that of benzene contaminated samples. Shin and Das [4] and Al-aghbari et al. [6] reported that oil with higher kinematic viscosity resulted in a higher reduction in permeability which is in agreement with the result of the present study.

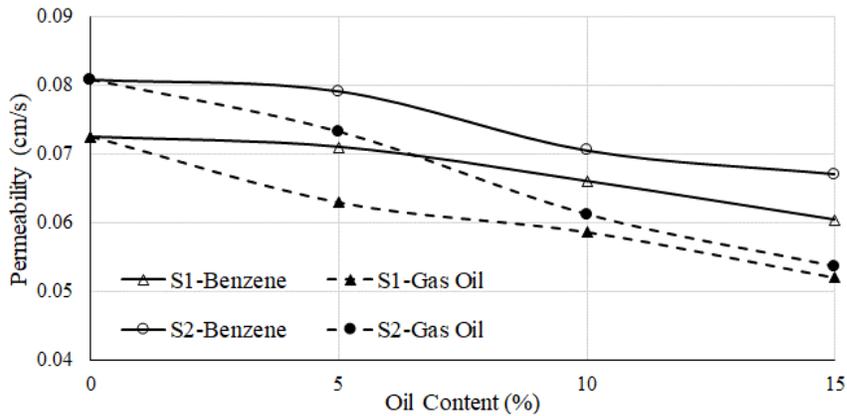

Figure 10. Effect of gas oil and benzene contamination on the permeability of S1 and S2 samples

## 4. Conclusion

Several geotechnical tests were carried out on clean and hydrocarbon-contaminated samples to investigate the influence of hydrocarbon contamination on two different types of sandy soils. The main findings, according to the results and discussions are as following:

1- According to standard Proctor test results, the bell-shaped compaction diagram for hydrocarbon-contaminated samples showed a tendency to the left side of the uncontaminated sample, which means that hydrocarbon-contaminated samples reached to their maximum dry density in lower moisture content.

2- Based on the results of the compaction tests, maximum dry densities increased up to 5% oil content. By increasing gas oil and benzene contents more than 5%, the maximum dry densities decreased due to the dissipation of compaction energy by excess fluid in the pore space.

3- The rate of changes in moisture content and maximum dry density for gas oil-contaminated samples was higher than that of benzene-contaminated samples which were attributed to the different viscosity of hydrocarbons.

4- The results of the direct shear tests demonstrated that even though cohesion increased in hydrocarbon-contaminated samples, the internal friction angle decreased.

5- The results of the direct shear tests revealed that the rate of changes in cohesion and internal friction angle for samples contaminated with gas oil was higher than that of benzene contaminated samples. The viscosity of gas oil is higher than the viscosity of benzene resulting in the more lubricating effect of gas oil than benzene.

6- The constant head permeability test showed that gas oil and benzene reduced the hydraulic permeability of samples. The rate of changes in hydraulic permeability was a function of oil content and viscosity.